\documentclass[USenglish,twocolumn]{article}

\ifx\directlua\undefined\ifx\XeTeXcharclass\undefined
  \usepackage[utf8]{inputenc}                           
  \else\RequirePackage[no-math]{fontspec}[2017/03/31]\fi 
  \else\RequirePackage[no-math]{fontspec}[2017/03/31]\fi 
\usepackage[sort&compress,square,numbers]{natbib}
\usepackage[big,online]{dgruyter}

\theoremstyle{dgthm}

\theoremstyle{dgdef}

\begin{document}

\articletype{Research Article}

\author[1]{Valerie Yoshioka}
\author[2]{Jicheng Jin}
\author[3]{Haiqi Zhou}
\author[4]{Zichen Tang}
\author[5]{Roy H. Olsson III}
\author*[6]{Bo Zhen}
\affil[1]{University of Pennsylvania, Philadelphia, Pennsylvania 19104, USA}
\affil[2]{University of Pennsylvania, Philadelphia, Pennsylvania 19104, USA}
\affil[3]{University of Pennsylvania, Philadelphia, Pennsylvania 19104, USA}
\affil[4]{University of Pennsylvania, Philadelphia, Pennsylvania 19104, USA}
\affil[5]{University of Pennsylvania, Philadelphia, Pennsylvania 19104, USA}
\affil[6]{University of Pennsylvania, Philadelphia, Pennsylvania 19104, USA, bozhen@sas.upenn.edu, https://orcid.org/0000-0001-8815-3893}

\title{CMOS-Compatible, AlScN-Based Integrated Electro-Optic Modulator}
\runningtitle{AlScN-Based EOM}
\abstract{Commercial production of integrated photonic devices is limited by scalability of desirable material platforms. We explore a relatively new photonic material, AlScN, for its use in electro-optic modulation. Its CMOS-compatibility could facilitate large-scale production of integrated photonic modulators, and it exhibits an enhanced second-order optical nonlinearity compared to intrinsic AlN, indicating the possibility for efficient modulation. Here, we measure the electro-optic effect in AlScN-based modulators, demonstrating $V_{\pi}L$ around 750 V$\cdot$cm. Since the electro-optic response is smaller than expected, we discuss potential causes for the reduced response and future outlook for AlScN-based photonics.}

\journalname{Nanophotonics}
\journalyear{2024}

\maketitle

\section{Introduction}
Integrated photonics promises control over light signals in small, chip-size packages, enabling signal processing with lower power consumption. An integral component in complex photonic devices is modulation, which allows dynamic control of light with a single chip. Information can be transferred between electrical and optical signals, acting as an interface between traditional electronic computers and low-loss optical fiber networks. While there are many methods to modulate light, ranging from thermo-optic to plasma dispersion, one of the most useful is electro-optic modulation based on the electro-optic (EO) or Pockels effect, as its fast intrinsic speed could allow terahertz bandwidth. Though experimental devices are still limited by factors like impedance mismatch and phase matching between RF and optical signals \cite{Kharel2021}, they regularly achieve bandwidth around 10s of GHz \cite{Sinatkas2021,Zhang2021,Renaud2023}, limited only by structural design. Well-designed modulators are capable of bandwidths exceeding 100 GHz \cite{Valdez2023, Mercante2018}. In addition to design, modulation efficiency relies on the strength of the electro-optic coefficient, which is material-dependent. However, well-developed photonic materials with strong electro-optic coefficients are limited. 

Silicon guides light on-chip based on its high refractive index contrast, which increases modal confinement and reduces footprint \cite{Petrov2021}. Its CMOS-compatibility facilitates fabrication, reducing costs. However, bulk silicon is centrosymmetric and thus cannot use the electro-optic effect for modulation. Silicon modulators are still possible via DC Kerr effect or plasma dispersion, but each have difficulties. For the DC Kerr effect, a large bias field is required to effectively enable electro-optic modulation, increasing power consumption \cite{Timurdogan2017}. Plasma dispersion modulators (PDMs) control free carrier density to adjust refractive index, resulting in two issues: 1) tuning index requires free carrier movement, introducing an intrinsic speed limit that greatly limits modulation bandwidth \cite{Deng2019} and 2) carrier density also impacts absorption, introducing unwanted intensity modulation and making higher-order modulation schemes more difficult \cite{Jacques2018}. Approaches to improve speed often reduce modulation efficiency \cite{Petrov2021}. Removing chirp requires digital post-processing \cite{Zhalehpour2020} or additional tuning with thermo-optic phase shifters, using more energy \cite{Deng2019}. 

In comparison, electro-optic modulation is intrisically fast and chirp-free, with efficiency only limited by material properties and design. Lithium niobate (LN) is a reliable optical material with low loss, large electro-optic coefficient ($r_{33} \sim$ 31 pm/V \cite{ZhuD2021}), and mature fabrication process. Its strong performance in electro-optic modulators (EOMs) has been well-documented \cite{Zhang2021}, including operation at reasonably low voltages \cite{Wang2018}. The main drawback in using LN is that its fabrication is not compatible with CMOS foundries, resulting in more complicated and expensive large-scale production. 

Aluminum nitride (AlN) is one of the few CMOS-compatible materials that exhibits an intrinsic electro-optic effect and sufficient photonic properties. Its electro-optic coefficient is lower than that of LN ($r_{13}$, $r_{33} \sim$ 1 pm/V \cite{Xiong2012}), but AlN can still be used in electro-optic modulation. However, a potential improvement in its performance could be controlled through the use of substitutional Sc atoms. 

Introducing Sc in AlN softens the crystal lattice, enhancing piezoelectric coefficients \cite{Akiyama2009} and second-order optical nonlinearity ($\chi^{(2)}$) \cite{Yoshioka2021}. The higher the Sc concentration, the larger the enhancement, until around 43$\%$ Sc. At this concentration, the crystal structure starts shifting from wurtzite to cubic structure, regaining centrosymmetry and eliminating these properties. While the electro-optic effect is often considered as a linear process since the shift in refractive index scales linearly with the applied electric field, it can be related to $\chi^{(2)}$ as it involves frequency mixing between a low-frequency applied electric field and the optical electric field \cite{Boyd}. Thus, enhancements in $\chi^{(2)}$ from frequency mixing between optical signals could indicate a larger electro-optic response. In this work, we demonstrate electro-optic modulation using Al$_{0.80}$Sc$_{0.20}$N. We utilized integrated Mach-Zehnder interferometer (MZI) devices to detect slight shifts in refractive index from applied voltage and to measure $V_{\pi}L$. 

\section{Theory}
To determine how an applied electric field affects refractive index in AlScN, we utilize the formulation for the electro-optic effect in wurtzite crystals \cite{Boyd}, which can be expressed by:
\begin{align}
    \Delta \left(\frac{1}{n^2}\right)_i &= \begin{bmatrix}
        0 & 0 &r_{13}\\
        0 &0 &r_{13}\\
        0 &0 &r_{33}\\
        0 &r_{51} &0\\
        r_{51} &0 &0\\
        0 &0& 0    \end{bmatrix}
         \begin{bmatrix}
             E_x\\
             E_y\\
             E_z
         \end{bmatrix}
\end{align}
By solving for refractive index, we can determine how an externally applied electric field shifts index. Due to in-plane polycrystallinity in our AlScN samples, the contributions from in-plane electric field components, $E_x$ and $E_y$, are expected to cancel on average due to opposing domain directions. As such, $r_{51}$ should not affect the overall response of our device, and only the $E_z$ component should result in a measurable electro-optic effect. With this simplification, the new refractive indices can be expressed as:
\begin{align}
    n_o' &= n_o - \frac{1}{2} n_o^3r_{13}E_{z}\\
    n_e' &= n_e - \frac{1}{2} n_e^3r_{33}E_{z}
\end{align}
Both ordinary and extraordinary indices are affected by an applied $E_{z}$ field, with $r_{13}$ controlling the change in $n_o$ and $r_{33}$ affecting $n_e$. 

In AlN, $r_{33}$ and $r_{13}$ are both around 1 pm/V \cite{Xiong2012}. To predict how EO coefficients in AlScN are enhanced, we can relate optical nonlinearity $\chi^{(2)}_{ijk} = 2 d_{ijk}$ to electro-optic coefficients \cite{Boyd1973}. Using full notation, we can relate the two values as:
\begin{align}
    r_{ijk} = \frac{-4d_{ijk}}{n_i^2n_j^2}
\end{align}
where $n_i$ is the refractive index along the $i$-axis. For AlN, $d_{33}$ is 5.1 pm/V \cite{Yoshioka2021}, corresponding to $r_{33}$ = 1.2 pm/V, which is similar to experimental values. For Al$_{0.80}$Sc$_{0.20}$N, we measured $d_{33}$ through second harmonic generation in the telecom regime to be 42.5 pm/V \cite{Yoshioka2021}. Assuming $d_{33}$ remains the same for the electro-optic effect as well, the corresponding $r_{33}$ would reach 8.1 pm/V, which is about a factor of 8 larger than intrinsic AlN. However, as we show later, this assumption may not be valid.

\section{Design}
\begin{figure*}[ht!]
    \centering
    \includegraphics[width=\textwidth]{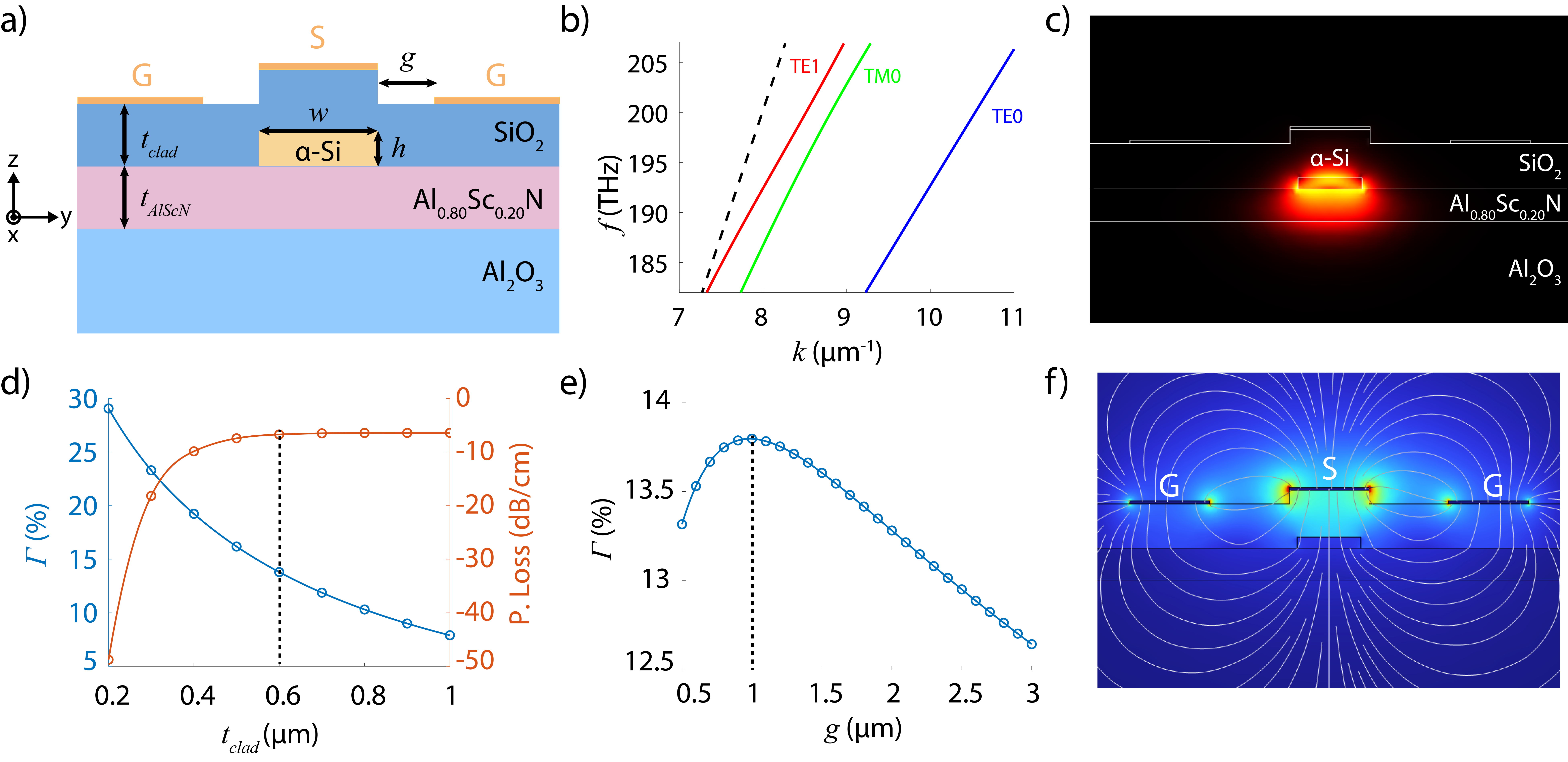}
    \caption{AlScN-based waveguide simulations. a) Cross-section of waveguide structure with labeled materials and dimensions. Electrodes are labeled with G for ground and S for signal. b) Modal dispersion for TE0, TM0, and TE1 modes for $w=800$ nm and $h=150$ nm; all modes are below the light line (dotted black line) and therefore guided modes. c) |E| for TM0 mode for $w=800$ nm and $h=150$ nm. A signficant portion of the modal field is present in the AlScN layer. d) Modal overlap $\Gamma$ and waveguide propagation loss as a function of oxide thickness $t_{clad}$; the final design used $t_{clad}$ = 600 nm. e) Modal overlap $\Gamma$ as a function of electrode gap $g$; the final design used $g$ = 1 $\mu$m. f) Electric field generated by applying voltage to the middle electrode in the GSG configuration; note that the field is primarily out-of-plane where the optical mode is located.}
    \label{Fig1}
\end{figure*}

Our platform is comprised of a sapphire substrate with $\left<0001\right>$ orientation, co-sputtered AlScN, an etched amorphous silicon ($\alpha$-Si) waveguide, PECVD SiO$_2$ cladding, and gold electrodes (Fig. 1a). Measured AlScN thickness was 429 nm. We designed the devices to use the fundamental TM0 mode, which has an electric field directed along the extraordinary axis. As such, it enables use of the $r_{33}$ electro-optic coefficient in AlScN, which is expected to be the larger coefficient. The $\alpha$-Si waveguide has $w$ = 800 nm and $h$ = 150 nm. We chose these dimensions to balance good confinement in the AlScN layer with reasonable loss in fabricated devices. Based on the dispersion, we confirmed that the TM0 mode is guided (Fig. 1b). Its electric field is also strongly localized in the AlScN layer despite the $\alpha$-Si waveguide being used as the index contrast for guiding light (Fig. 1c). While there is slight hybridization between the TM0 mode and TE1 mode at a waveguide width of 800 nm, we found that using a slightly wider waveguide reduced loss in fabricated devices. Since the mode is weakly confined in the etched silicon waveguide, a large bend radius around 250 $\mu$m was necessary to reduce bend loss.

We used coplanar electrodes in a ground-signal-ground (GSG) configuration to generate an out-of-plane electric field \cite{Xiong2012}. In order to generate a strong electro-optic response, the modal overlap between the applied electric field and optical mode needs to be large to encourage interaction between the fields. We define modal overlap as:
\begin{align}
    \Gamma = \frac{g}{V_{in}} * \frac{\iint (E_{z,op}^2 E_z) dx dz}{\iint (E_{z,op}^2) dx dz}
\end{align}
where $E_{z,op}$ is the out-of-plane electric field of the TM0 optical mode, $E_z$ is the applied out-of-plane electric field, $V_{in}$ is voltage applied to the signal electrode, and $g$ is the horizontal gap between electrodes. The integrals are evaluated in the region of the electro-optic material. In order to maximize modal overlap, we optimized oxide thickness and the horizontal gap between the electrodes. Oxide thickness $t_{clad}$ was set to be 600 nm to maximize effective electro-optic response, while keeping mode loss reasonable (Fig. 1d). Electrode gap $g$ was set to 1 $\mu$m to maximize modal overlap between the applied field and optical mode (Fig. 1e). As a result, we can generate a strong out-of-plane field in the region of the waveguide, with reasonable modal overlap around 14$\%$ (Fig. 1f).

In order to couple light into and out of the integrated photonic device, we designed grating couplers with focusing geometry. We used grating period $a$ = 0.87 $\mu$m, fill factor 60$\%$, and focusing angle of 40° to couple light into a fiber array with polish angle of 8°. By using grating couplers, we ensure mode selectivity as the grating period is tuned to only couple in light with the desired effective mode index. 

\section{Fabrication}
Al$_{0.80}$Sc$_{0.20}$N was deposited onto the sapphire substrate via co-sputtering in a pure nitrogen environment using a pulsed DC physical vapor deposition system (Evatec CLUSTERLINE® 200 II). Relative power applied to the Al and Sc targets was adjusted to control relative Sc concentration in the films. A thin seed layer was grown using 250 W on the Sc target and 875 W on the Al target to facilitate lattice matching to the substrate and ensure well-ordered crystal structure. Sc concentration was then linearly graded to achieve 20$\%$ Sc concentration in the bulk layer. While polycrystalline in-plane, the c-axis is well-oriented perpendicular to the substrate. Its ordinary refractive index along the in-plane directions was measured to be 2.124 via prism coupling at 1550 nm, which is consistent with other refractive index measurements \cite{Wingqvist2010, Baeumler2019}. Extraordinary index is oriented along the c-axis and is expected to be slightly larger, around 2.14 \cite{Wingqvist2010}. Loss at 1550 nm was measured via prism coupling to be 8.67 dB/cm, which is consistent with our prior measurements of similar samples \cite{Yoshioka2021}. AFM measurements indicate reasonable roughness, with $R_q$ = 4.19 nm and $R_a$ = 3.48 nm over a large area of 5 $\times$ 5 $\mu$m$^2$. 

To make devices, intrinsic amorphous silicon ($\alpha$-Si) was deposited on top of the AlScN layer via RF sputtering (Denton Explorer14 Magnetron Sputterer). Device patterns were defined using e-beam lithography (EBL). The e-beam resist (ZEP520A-07) was chemically developed using O-Xylene in a cold bath around -5°C to -10°C in order to reduce sidewall roughness. The $\alpha$-Si was subsequently etched using CF$_4$ in a dry reactive ion etching process (Oxford 80 Plus). After stripping the remaining resist using NMP in a heated ultrasonic bath, oxide was grown via PECVD as the top cladding layer, and the passive photonic response was measured. The thickness of each layer was confirmed by cross-sectional SEM (Fig. 2a).

Electrode patterns were defined via EBL using PMMA resist. Square markers from the first EBL exposure enabled sufficient alignment between electrode and waveguide patterns. Once the electrode pattern was developed, we deposited Ti/Au via e-beam evaporation; a thin 5 nm layer of Ti was used to adhere the 40 nm layer of gold to the surface oxide in order to form the electrodes. Heated NMP was used to remove the remaining metal and resist, leaving the electrodes on top. Microscope images confirm good alignment between the deposited electrodes and underlying waveguides (Fig. 2b). 

To determine insertion loss, we fabricated test devices comprised of two grating couplers connected by a short waveguide (Fig. 2c). The total insertion loss of a test device was around -31 dB (Fig. 2d), which corresponds to insertion loss around -15 dB per coupler. Though insertion loss could be improved through additional techniques such as apodization and shallow etching, this straightforward design is easier to fabricate and achieves sufficiently low loss in order to perform electro-optic measurements. For propagation loss, we measured devices with different waveguide lengths and found loss was around 10 $\pm$ 2 dB/cm (Fig. 2e). Since the intrinsic loss of Al$_{0.80}$Sc$_{0.20}$N was measured to be almost 9 dB/cm, our propagation loss is likely limited by intrinsic loss rather than scattering or sidewall roughness. 

\begin{figure}[ht!]
    \centering
    \includegraphics[width=8cm]{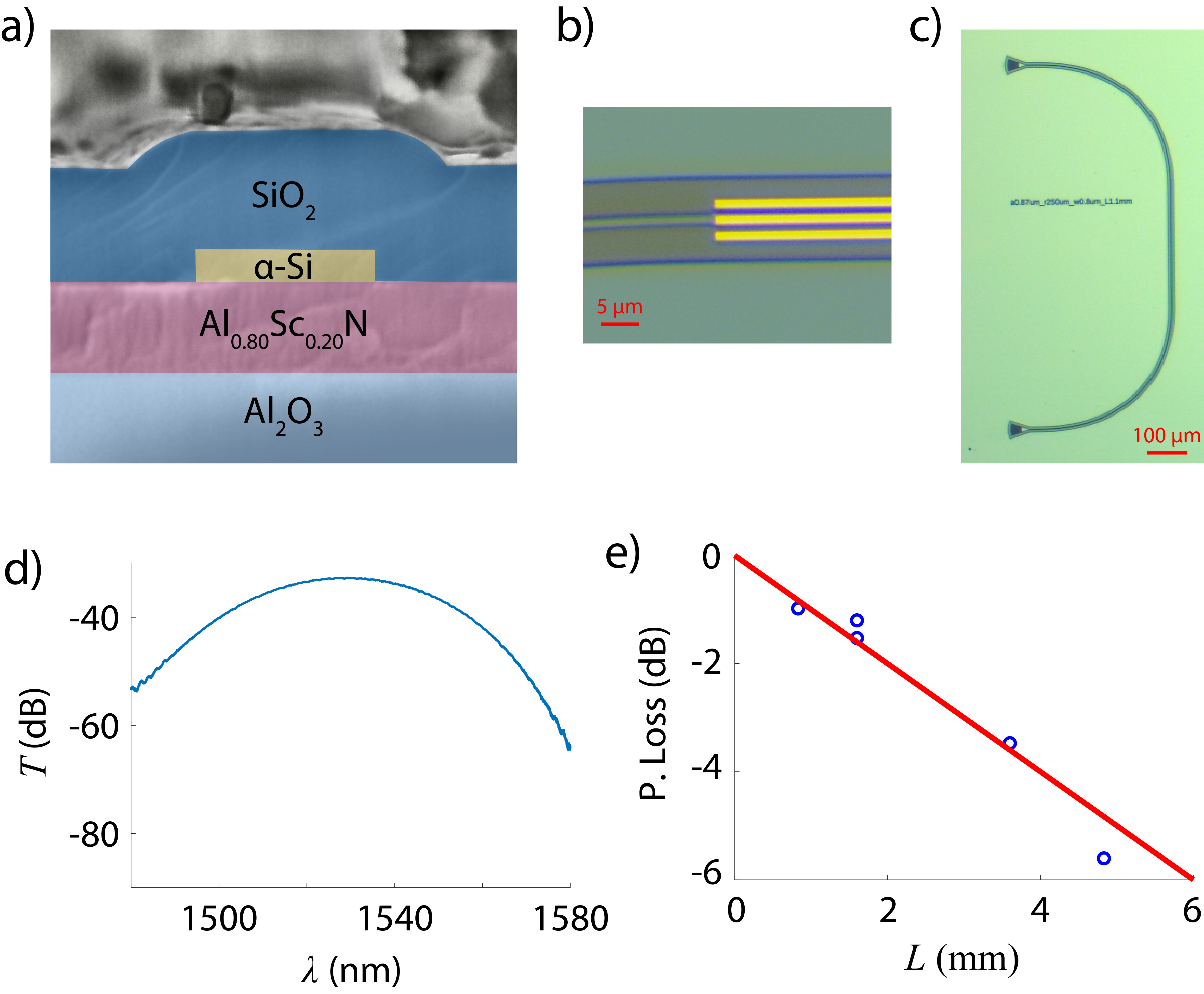}
    \caption{Fabricated AlScN-based photonic devices. a) Cross-sectional SEM image of waveguide confirms expected dimensions. b) Microscope image of electrodes on top of a waveguide, visually confirming good alignment. c) Microscope image of an AlScN-based photonic device with two grating couplers connected by a short waveguide. d) Transmission for a test device; insertion loss from a single grating coupler is around -15 dB. e) Propagation loss for the waveguides is around 10 $\pm$ 2 dB/cm.}
    \label{Fig2}
\end{figure}
 
\section{Results}
To measure our devices, we used a telecom laser (Keysight 8164B) as the light source, connected to a 3-paddle fiber polarization controller. The fiber was then connected to an input on a fiber array in order to couple light onto the chip via grating couplers. The output light signal was directed to a power meter (Keysight N7744A) to determine transmission as a function of wavelength. We adjusted the chip position, chip rotation, and input polarization to maximize transmission for test devices of two grating couplers connected by a short waveguide. Once the positioning was optimized, we proceeded to measure more complex devices on the chip. 

In order to detect a small index change induced by the electro-optic effect, we utilized Mach-Zehnder interferometers (MZIs). These devices work by splitting the input light into two branches, applying voltage to electrodes above one waveguide branch, and recombining the branches into a single output signal. Depending on wavelength, the light experiences either constructive or destructive interference, resulting in a pattern of fringes when wavelength is swept. The transmission, $T = \frac{I_{out}}{I_{in}}$, at applied voltage $V_{in}$ can be expressed as: 
\begin{align}
    T &= \frac{T_{max}}{2}\left[1+cos\left(\frac{2 \pi}{\lambda}(n_{\rm eff}\Delta L - \Delta n_{\rm eff} L_1)\right)\right]
\end{align}
where $L_{1,2}$ are the path lengths for respective branches, with $L_2 = L_1 + \Delta L$, $\lambda$ is the guided wavelength, $T_{max}$ is the maximum amplitude of the transmission based on total loss, $n_{\rm eff}$ is the effective index for the TM0 mode, and $\Delta n_{\rm eff}$ is the index change induced in the $L_1$ branch by voltage $V_{in}$. We utilized a path length difference of 250 $\mu$m for our devices, resulting in a free spectral range around 4.2 nm (Fig. 3a). The extinction ratio in the fringes varies between -20 to -30 dB, which is large enough to facilitate electro-optic measurements. Using the MZI transmission equation, we can solve for $\Delta n_{\rm eff}$ by measuring the change in transmission, $\Delta T = T(0)-T(V_{in})$. Using the small angle approximation and the wavelength $\lambda_0$ at which $\Delta T$ is maximized, we can express $\Delta n_{\rm eff}$ as:
\begin{align}
    |\Delta n_{\rm eff}| &= \frac{\lambda_{0} \Delta T}{\pi L_1 T_{max}}
\end{align}
$\Delta n_{\rm eff}$ can then be related to $V_\pi L$ by:
\begin{align}
    V_\pi L &= \frac{V_{in} \lambda_0}{2 \Delta n_{\rm eff}}
\end{align}

To determine our device's performance, we applied DC voltage to the electrodes using an RF probe, functional from DC to 40 GHz, in direct contact with the chip. The probe was connected to a DC power supply to apply voltage. We measured the optical responses $T(V_{in} = 0V)$ and $T(V_{in} = 10V)$, observing a shift between the two due to the electro-optic effect (Fig. 3b). The induced shift in transmission, $\Delta T$, is maximized at the largest slope of the transmission signal, as expected. $\Delta T$ is about 400 times smaller than the maximum transmission, indicating a small electro-optic response. We confirmed that this response was due to the electro-optic effect by sweeping input voltage and measuring a linear trend in the response (Fig. 3c). Due to the small signal, we used an InGaAs photodetector (Thorlabs PDA10DT) connected to a lock-in amplifier (Stanford Research Systems SR865A) to confirm this linear trend. Due to differences between the photodetector and power meter, $\Delta T$ was measured in volts using the lock-in amplifier and watts for DC measurements. 

\begin{figure*}[ht!]
    \centering
    \includegraphics[width=\textwidth]{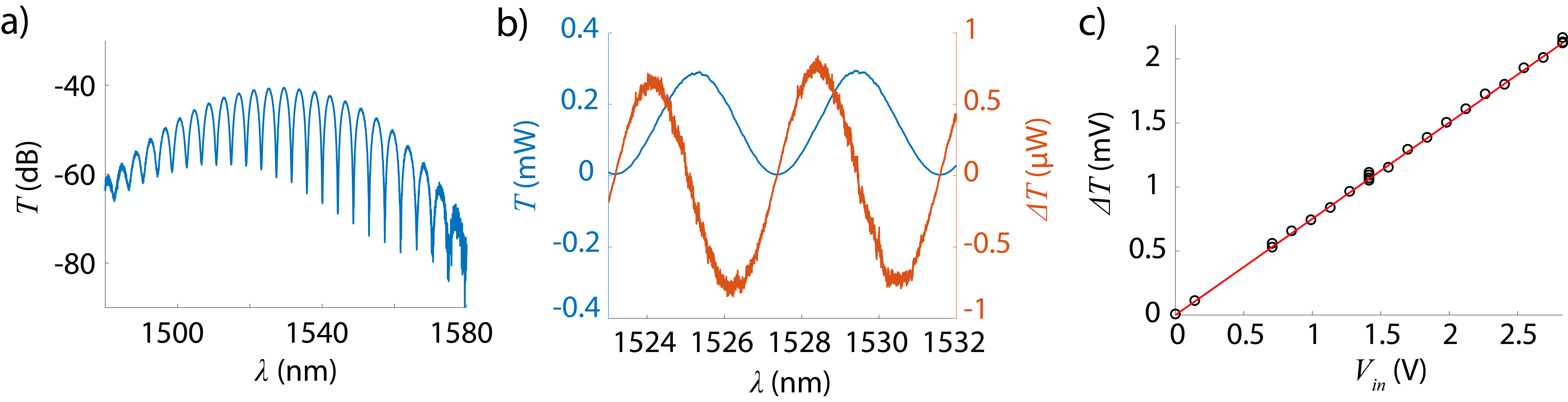}
    \caption{Electro-optic response of AlScN device. a) Transmission for a Mach-Zehnder interferometer with $\Delta L$ = 250 $\mu$m; free spectral range is around 4.2 nm. b) Transmission $T$ and change in transmission $\Delta T$ for an AlScN-based electro-optic modulator on a linear scale. $\Delta T$ reaches a maximum when $T$ experiences its largest slope. c) Linear increase in $\Delta T$ versus applied voltage $V_{in}$, as measured by lock-in amplifier.}
    \label{Fig3}
\end{figure*}

Based on DC measurements, we found our device experienced $\Delta n_{\rm eff} = 8.2 \times 10^{-8}$ per volt, which corresponds to a 3.4 pm wavelength shift after applying 10 V and $V_\pi L = 933$ V$\cdot$cm. Using more precise lock-in measurements with a driving frequency of 200 Hz, we found a slightly larger response of $\Delta n_{\rm eff} = 1 \times 10^{-7}$ per volt, corresponding to $V_\pi L = 750$ V$\cdot$cm. These responses are within 20$\%$ of each other, with a smaller DC response. We also measured a second device at 100 Hz and found a similar response of $\Delta n_{\rm eff} = 9.6 \times 10^{-8}$ per volt, corresponding to $V_\pi L = 800$ V$\cdot$cm. In lock-in amplifier measurements, we primarily measured the EO effect around 100-200 Hz but found little variation in EO response between 1 Hz and 10 kHz. 

The difference between the DC and lock-in measurements could be related to EO relaxation, which has been observed in low-frequency measurements of LN-based EOMs \cite{Zhang2021} and can be broadly attributed to free-carrier charges moving in the material \cite{Holzgrafe2023}. For LN, where the effect is strongly present, conductivity is around $1 \times 10^{-6}$ S/m. While AlN has very low conductivity around $1.1 \times 10^{-12}$ S/m \cite{Kim2015}, the introduction of Sc in AlScN can increase conductivity to around $1 \times 10^{-7}$ to $1 \times 10^{-5}$ S/m \cite{Zheng2023}. There are methods to reduce the relaxation effect such as removing top cladding \cite{Xu2021} or performing measurements at sufficiently high frequencies \cite{Zhang2021}, but material interfaces, defects, and fabrication parameters can greatly complicate predictions of this effect \cite{Holzgrafe2023}. However, it's unclear how much the Sc concentration affects EO relaxation on different timescales. 

\section{Discussion}
Based on our simulations, we expected $\Delta n_{\rm eff}$ to be $2.8 \times 10^{-6}/V$ for Al$_{0.80}$Sc$_{0.20}$N based on its enhanced $\chi^{(2)}$ or $\Delta n_{\rm eff} = 4.4 \times 10^{-7}/V$ assuming $r_{33}$, $r_{13}$ = 1 pm/V. An MZI-based electro-optic modulator with AlN reported $\Delta n_{\rm eff} = 2.4 \times 10^{-7}/V$ for the TM mode \cite{Zhu2016}, which is similar to our expected performance for intrinsic AlN. However, our fabricated devices have low EO responses around $\Delta n_{\rm eff} = 1 \times 10^{-7}/V$, which are more similar to AlN and do not demonstrate any enhancement. There are a few possible explanations. 

Some of the reduced performance comparing simulations to measurements is due to imperfections from fabrication. The overlap between electric field and mode could be smaller than expected due to slight variation in waveguide parameters or material properties. Electrode alignment also plays a factor in total measured signal. Visually, the electrode is aligned within 1 $\mu$m of the waveguide, but even slight misalignments can reduce EO response. Invisible factors that cause uneven application of field energy, such as variation in the silicon or oxide thickness, roughness on the oxide surface, and imperfections in the electrodes, can similarly reduce EO response. Note that the DC permittivity of AlScN also changes with concentration. For intrinsic AlN, $\epsilon_r$ is around 9.9, while for Al$_{0.80}$Sc$_{0.20}$N, $\epsilon_r$ increases to 13.7 \cite{Wingqvist2010}. Higher permittivity reduces electric field in the material, slightly reducing overall efficiency for EO modulation. However, these factors would have a relatively small effect and are not sufficiently large to indicate enhancement in the electro-optic coefficients. 

In terms of sample quality, a film of surface oxidation on the AlScN layer could reduce overall response, particularly since the optical mode is concentrated at the top surface of the AlScN. Surface oxidation would therefore reduce the effective mode overlap and the electro-optic response as a result. There is also evidence that oxidation can extend into the bulk of the sputtered AlScN when the crystal structure exhibits many grain boundaries \cite{Li2022}. Since oxidation depends on film structure, its extent varies depending on growth method and sample quality. While it's unclear how much surface oxidation affects electro-optic response in our sample, methods to prevent or remove this oxidation would likely improve performance in future devices. Fully etched waveguide designs may also have better performance as they allow mode overlap to be localized further away from oxidized surface areas.

Another explanation for the reduced response could be due to varying contributions from the EO tensor for AlScN. Based on crystal symmetry and our sample's in-plane polycrystallinity, we expect the $r_{51}$ component does not contribute to overall response for our geometry, while $r_{33}$ and $r_{31}$ can both contribute due to an applied out-of-plane electric field. While our device uses the fundamental TM mode in the optical waveguide, allowing $r_{33}$ response, the slight hybridization between TM0 and TE1 modes also allows $r_{31}$ to contribute. Based on simulations, $r_{33}$ should still dominate the response. However, the relative strength and signs of $r_{33}$ and $r_{31}$ are difficult to extract from this measurement. The introduction of Sc could affect each EO component in ways that do not necessarily match measured $\chi^{(2)}$ enhancements, especially since phonon resonances exist between the mixed frequencies in the electro-optic effect. Different EO components could also interact in such a way as to reduce the overall EO effect even if individual components are enhanced, particularly if the coefficients have opposite signs. In AlN, there is some evidence that EO components have different signs \cite{Graupner1992}, though measured values and relative signs are somewhat contradictory \cite{Majkic2017}. Future research should seek to systematically measure changes in each of these components separately for different Sc concentrations. 

It is also possible that the electro-optic coefficients are not enhanced like $\chi^{(2)}$ due to significant material dispersion and phonon resonances between the mixed frequencies. The electro-optic effect has contributions from ionic terms in addition to $\chi^{(2)}$ \cite{Veithen2004}, which could diminish the EO effect. Theoretical studies of AlScN crystal structure for different Sc concentrations would shed light on whether ionic contributions are a limiting factor in EO response and how these ionic terms depend on Sc concentration. It's also possible that the EO effect is enhanced for some Sc concentrations but not others. 

Furthermore, piezoelectricity can also affect electro-optic response at sufficiently low frequencies. In the DC regime, the applied electric field is well below acoustic resonances in the material, inducing strain due to the piezoelectric effect \cite{Sutherland}. As a result, the electro-optic response can be affected by the piezoelectric behavior. In our device, the AlScN in our devices is clamped between a substrate and surface cladding layer, so while it may experience some strain, it is unlikely that piezoelectric terms have a large effect even at low frequency. As such, it seems more likely that ionic terms could be responsible for the diminished EO effect in our geometry. However, in alternate geometries where the AlScN is free to move mechanically, it may be possible to use piezoelectricity to enhance the electro-optic effect at sufficiently low frequencies. 

\section{Future Outlook}
While this work demonstrates the difficulty in using AlScN for electro-optic modulation, modern fabrication advances could improve response, assuming the material itself is not the limiting factor. Previously, AlScN etching recipes were primarily limited by sidewall roughness \cite{Zhu2020}. To avoid this loss, we used strip-loaded silicon waveguides to guide light, as silicon etching recipes are better established and more reliable. While these waveguides still allow some interaction with the AlScN film underneath, the modal confinement in AlScN is limited. However, improvements in etching recipes over the past few years have enabled directly etched AlScN waveguides with smoother sidewalls \cite{Wang2024, Friedman2024}. As a result, light can be more strongly confined in AlScN to improve nonlinear response without concerns about sidewall roughness or surface oxidation. 

Intrinsic material loss, due to higher absorption with Sc concentration and polycrystalline structure, remains a significant limitation. However, the concentration of Sc can be controlled depending on the tolerance for loss. As for crystalline quality, there are methods for epitaxial growth of single crystal AlScN films, which could reduce loss regardless of Sc concentration \cite{Wang2020}. Techniques like polishing and annealing have also been demonstrated to reduce propagation loss \cite{Wang2024}. Much lower waveguide losses have been recently reported, indicating waveguide losses below 2 dB/cm in etched Al$_{0.70}$Sc$_{0.30}$N rib waveguides \cite{Friedman2024}. While their data suggests lower intrinsic loss, potentially due to their growth technique, their waveguides are also quite thin, reducing confinement in lossy AlScN. By adjusting waveguide design, confinement in AlScN could be reduced in passive guiding areas of a device to decrease loss and then increased in nonlinear or electro-optic areas to improve efficiency. Additional studies on growth of low-loss AlScN films would greatly improve future performance for AlScN-based photonics.

Additional flexibility in growing AlScN on different substrates could also improve overall response. In our devices, we used sapphire as a substrate to ensure highly oriented AlScN films due to lattice matching considerations, but as a result, we were unable to fabricate devices with vertical electrodes and used a less efficient coplanar design instead. Recent studies have used a thin seed layer of AlN to grow AlScN on oxide \cite{Su2022}, which is easier to work with and can be deposited on a variety of substrates. It would also allow the possibility of depositing a bottom electrode beneath the lower oxide layer for vertical electrodes, improving overlap between optical and electrical fields and thus overall response. 

Finally, periodic poling in AlScN could allow quasi-phase-matching for frequency mixing applications. Demonstrations of poling in Al$_{0.68}$Sc$_{0.32}$N have achieved poling widths as narrow as 250 nm \cite{Tang2023}. While poling in AlN has been demonstrated at higher temperatures \cite{ZhuW2021}, it is a relatively recent implementation and has not yet been widely utilized in integrated photonic devices. Thus, the introduction of Sc could facilitate room temperature poling to improve efficiency in CMOS-compatible, nonlinear integrated devices. 

\section{Conclusion}
We designed, fabricated, and measured electro-optic modulators based on AlScN. Modulators with Al$_{0.80}$Sc$_{0.20}$N had a measured performance of $V_{\pi}L$ = 750 V$\cdot$cm. We expect developments in AlScN fabrication techniques and modulator design have the potential to improve future AlScN-based modulators. While theoretical studies are needed to shed light on the intrinsic limits of AlScN, from its loss to its electro-optic response, its CMOS-compatibility and enhanced optical nonlinearity could still facilitate large scale production of other nonlinear integrated photonic devices. 

\begin{acknowledgement}
Zhi Wang provided input on fabrication methods. Fabrication and material characterization was performed at the Singh Center for Nanotechnology at the University of Pennsylvania, which is supported by the NSF National Nanotechnology Coordinated Infrastructure Program under grant NNCI-2025608. Prism coupling measurements were performed by Metricon Corporation.
\end{acknowledgement}

\begin{funding}
This work was funded by the Army Research Office (W911NF-19-1-0087) and the NSF CAREER Award (1944248). V.Y. acknowledges support from the Department of Defense National Defense Science and Engineering Graduate (NDSEG) Fellowship.
\end{funding}

\begin{authorcontributions}
All authors have accepted responsibility for the entire content of this manuscript and approved its submission.
\end{authorcontributions}

\begin{conflictofinterest}
Authors state no conflict of interest.
\end{conflictofinterest}

\begin{dataavailabilitystatement}
The datasets generated during and/or analyzed during the current study are available from the corresponding author on reasonable request. 
\end{dataavailabilitystatement}

\bibliographystyle{ieeetr}
\bibliography{sample}

\begin{thebibliography}{10}

\bibitem{Kharel2021}
P.~Kharel, C.~Reimer, K.~Luke, L.~He, and M.~Zhang, ``{Breaking voltage--bandwidth limits in integrated lithium niobate modulators using micro-structured electrodes},'' {\em Optica}, vol.~8, no.~3, pp.~357--363, 2021.

\bibitem{Sinatkas2021}
G.~Sinatkas, T.~Christopoulos, O.~Tsilipakos, and E.~E. Kriezis, ``{Electro-optic modulation in integrated photonics},'' {\em Journal of Applied Physics}, vol.~130, no.~1, 2021.

\bibitem{Zhang2021}
M.~Zhang, C.~Wang, P.~Kharel, D.~Zhu, and M.~Lon{\v{c}}ar, ``{Integrated lithium niobate electro-optic modulators: when performance meets scalability},'' {\em Optica}, vol.~8, no.~5, p.~652, 2021.

\bibitem{Renaud2023}
D.~Renaud, D.~R. Assumpcao, G.~Joe, A.~Shams-Ansari, D.~Zhu, Y.~Hu, N.~Sinclair, and M.~Lon{\v{c}}ar, ``{Sub-1 Volt and high-bandwidth visible to near-infrared electro-optic modulators},'' {\em Nature Communications}, vol.~14, no.~1, pp.~1--7, 2023.

\bibitem{Valdez2023}
F.~Valdez, V.~Mere, and S.~Mookherjea, ``{100 GHz bandwidth, 1 volt integrated electro-optic Mach–Zehnder modulator at near-IR wavelengths},'' {\em Optica}, vol.~10, no.~5, 2023.

\bibitem{Mercante2018}
A.~J. Mercante, S.~Shi, P.~Yao, L.~Xie, R.~M. Weikle, and D.~W. Prather, ``{Thin film lithium niobate electro-optic modulator with terahertz operating bandwidth},'' {\em Opt. Express}, vol.~26, no.~11, pp.~14810--14816, 2018.

\bibitem{Petrov2021}
V.~M. Petrov, P.~M. Agruzov, V.~V. Lebedev, I.~V. Il'ichev, and A.~V. Shamray, ``{Broadband integrated optical modulators: achievements and prospects},'' {\em Physics-Uspekhi}, vol.~64, no.~7, pp.~722--739, 2021.

\bibitem{Timurdogan2017}
E.~Timurdogan, C.~V. Poulton, M.~J. Byrd, and M.~R. Watts, ``{Electric field-induced second-order nonlinear optical effects in silicon waveguides},'' {\em Nature Photonics}, vol.~11, no.~3, pp.~200--206, 2017.

\bibitem{Deng2019}
H.~Deng and W.~Bogaerts, ``{Pure phase modulation based on a silicon plasma dispersion modulator},'' {\em Optics Express}, vol.~27, no.~19, p.~27191, 2019.

\bibitem{Jacques2018}
M.~Jacques, A.~Samani, D.~Patel, E.~El-Fiky, M.~Morsy-Osman, T.~Hoang, M.~G. Saber, L.~Xu, J.~Sonkoly, M.~Ayliffe, and D.~V. Plant, ``{Modulator material impact on chirp, DSP, and performance in coherent digital links: comparison of the lithium niobate, indium phosphide, and silicon platforms},'' {\em Opt. Express}, vol.~26, no.~17, pp.~22471--22490, 2018.

\bibitem{Zhalehpour2020}
S.~Zhalehpour, M.~Guo, J.~Lin, Z.~Zhang, Y.~Qiao, W.~Shi, and L.~A. Rusch, ``{System Optimization of an All-Silicon IQ Modulator: Achieving 100-Gbaud Dual-Polarization 32QAM},'' {\em Journal of Lightwave Technology}, vol.~38, no.~2, pp.~256--264, 2020.

\bibitem{ZhuD2021}
D.~Zhu, L.~Shao, M.~Yu, R.~Cheng, B.~Desiatov, C.~J. Xin, Y.~Hu, J.~Holzgrafe, S.~Ghosh, A.~Shams-Ansari, E.~Puma, N.~Sinclair, C.~Reimer, M.~Zhang, and M.~Lon{\v{c}}ar, ``{Integrated photonics on thin-film lithium niobate},'' {\em Advances in Optics and Photonics}, vol.~13, no.~2, p.~242, 2021.

\bibitem{Wang2018}
C.~Wang, M.~Zhang, X.~Chen, M.~Bertrand, A.~Shams-Ansari, S.~Chandrasekhar, P.~Winzer, and M.~Lon{\v{c}}ar, ``{Integrated lithium niobate electro-optic modulators operating at CMOS-compatible voltages},'' {\em Nature}, vol.~562, no.~7725, pp.~101--104, 2018.

\bibitem{Xiong2012}
C.~Xiong, W.~H.~P. Pernice, and H.~X. Tang, ``{Low-Loss, Silicon Integrated, Aluminum Nitride Photonic Circuits and Their Use for Electro-Optic Signal Processing},'' {\em Nano Letters}, vol.~12, no.~7, pp.~3562--3568, 2012.

\bibitem{Akiyama2009}
M.~Akiyama, T.~Kamohara, K.~Kano, A.~Teshigahara, Y.~Takeuchi, and N.~Kawahara, ``Enhancement of piezoelectric response in scandium aluminum nitride alloy thin films prepared by dual reactive cosputtering,'' {\em Advanced Materials}, vol.~21, no.~5, pp.~593--596, 2009.

\bibitem{Yoshioka2021}
V.~Yoshioka, J.~Lu, Z.~Tang, J.~Jin, R.~H. {Olsson III}, and B.~Zhen, ``{Strongly enhanced second-order optical nonlinearity in CMOS-compatible Al$_{1-x}$Sc$_x$N thin films},'' {\em APL Materials}, vol.~9, no.~10, p.~101104, 2021.

\bibitem{Boyd}
R.~W. Boyd, {\em Nonlinear Optics}.
\newblock Academic Press, third~ed., 1992.

\bibitem{Boyd1973}
G.~D. Boyd and M.~A. Pollack, ``{Microwave nonlinearities in anisotropic dielectrics and their relation to optical and electro-optical nonlinearities},'' {\em Physical Review B}, vol.~7, no.~12, pp.~5345--5359, 1973.

\bibitem{Wingqvist2010}
G.~Wingqvist, F.~Tasn{\'{a}}di, A.~Zukauskaite, J.~Birch, H.~Arwin, and L.~Hultman, ``{Increased electromechanical coupling in w – Sc$_x$Al$_{1-x}$N},'' {\em Applied Physics Letters}, vol.~97, no.~11, pp.~1--4, 2010.

\bibitem{Baeumler2019}
M.~Baeumler, Y.~Lu, N.~Kurz, L.~Kirste, M.~Prescher, T.~Christoph, J.~Wagner, A.~{\v{Z}}ukauskaitė, and O.~Ambacher, ``{Optical constants and band gap of wurtzite Al$_{1-x}$Sc$_{x}$N/Al$_2$O$_3$ prepared by magnetron sputter epitaxy for scandium concentrations up to x = 0.41},'' {\em Journal of Applied Physics}, vol.~126, no.~4, p.~45715, 2019.

\bibitem{Holzgrafe2023}
J.~Holzgrafe, E.~Puma, R.~Cheng, H.~Warner, A.~Shams-Ansari, R.~Shankar, and M.~Lon{\v{c}}ar, ``{Relaxation of the electro-optic response in thin-film lithium niobate modulators},'' {\em Optics Express}, vol.~32, no.~3, pp.~16--19, 2023.

\bibitem{Kim2015}
T.~Kim, J.~Kim, R.~Dalmau, R.~Schlesser, E.~Preble, and X.~Jiang, ``{High-temperature electromechanical characterization of AlN single crystals},'' {\em IEEE Transactions on Ultrasonics, Ferroelectrics, and Frequency Control}, vol.~62, no.~10, pp.~1880--1887, 2015.

\bibitem{Zheng2023}
J.~X. Zheng, M.~M.~A. Fiagbenu, G.~Esteves, P.~Musavigharavi, A.~Gunda, D.~Jariwala, E.~A. Stach, and R.~H. {Olsson III}, ``{Ferroelectric behavior of sputter deposited Al$_{0.72}$Sc$_{0.28}$N approaching 5 nm thickness},'' {\em Applied Physics Letters}, vol.~122, no.~22, p.~222901, 2023.

\bibitem{Xu2021}
Y.~Xu, M.~Shen, J.~Lu, J.~B. Surya, A.~A. Sayem, and H.~X. Tang, ``{Mitigating photorefractive effect in thin-film lithium niobate microring resonators},'' {\em Optics Express}, vol.~29, no.~4, p.~5497, 2021.

\bibitem{Zhu2016}
S.~Zhu and G.-Q. Lo, ``{Aluminum nitride electro-optic phase shifter for backend integration on silicon},'' {\em Opt. Express}, vol.~24, no.~12, pp.~12501--12506, 2016.

\bibitem{Li2022}
M.~Li, K.~Hu, H.~Lin, V.~Felmetsger, and Y.~Zhu, ``{Oxidation of sputtered AlScN films exposed to the atmosphere},'' {\em IEEE International Ultrasonics Symposium, IUS}, vol.~2022-October, pp.~1--3, 2022.

\bibitem{Graupner1992}
P.~Gräupner, J.~C. Pommier, A.~Cachard, and J.~L. Coutaz, ``{Electro‐optical effect in aluminum nitride waveguides},'' {\em Journal of Applied Physics}, vol.~71, no.~9, pp.~4136--4139, 1992.

\bibitem{Majkic2017}
A.~Majkić, A.~Franke, R.~Kirste, R.~Schlesser, R.~Collazo, Z.~Sitar, and M.~Zgonik, ``Optical nonlinear and electro-optical coefficients in bulk aluminium nitride single crystals,'' {\em physica status solidi (b)}, vol.~254, no.~9, p.~1700077, 2017.

\bibitem{Veithen2004}
M.~Veithen, X.~Gonze, and P.~Ghosez, ``{First-principles study of the electro-optic effect in ferroelectric oxides},'' {\em Physical Review Letters}, vol.~93, no.~18, pp.~1--4, 2004.

\bibitem{Sutherland}
R.~L. Sutherland, {\em Handbook of Nonlinear Optics}.
\newblock Marcel Dekker, Inc., second~ed., 2003.

\bibitem{Zhu2020}
S.~Zhu, Q.~Zhong, N.~Li, T.~Hu, Y.~Dong, Z.~Xu, Y.~Zhou, Y.~H. Fu, and N.~Singh, ``{Integrated ScAlN Photonic Circuits on Silicon Substrate},'' in {\em Conference on Lasers and Electro-Optics}, p.~STu3P.5, Optica Publishing Group, 2020.

\bibitem{Wang2024}
S.~Wang, V.~Dhyani, S.~S. Mohanraj, X.~Shi, B.~Varghese, W.~W. Chung, D.~Huang, Z.~S. Lim, Q.~Zeng, H.~Liu, X.~Luo, V.~Leong, N.~Li, and D.~Zhu, ``{CMOS-compatible photonic integrated circuits on thin-film ScAlN},'' pp.~2--7, 2024.

\bibitem{Friedman2024}
B.~Friedman, S.~Barth, T.~Schreiber, H.~Bartzsch, J.~Bain, and G.~Piazza, ``{Measured optical losses of Sc doped AlN waveguides},'' {\em Opt. Express}, vol.~32, no.~4, pp.~5252--5260, 2024.

\bibitem{Wang2020}
P.~Wang, D.~A. Laleyan, A.~Pandey, Y.~Sun, and Z.~Mi, ``{Molecular beam epitaxy and characterization of wurtzite Sc$_x$Al$_{1-x}$N},'' {\em Applied Physics Letters}, vol.~116, no.~15, 2020.

\bibitem{Su2022}
J.~Su, S.~Fichtner, M.~Z. Ghori, N.~Wolff, M.~R. Islam, A.~Lotnyk, D.~Kaden, F.~Niekiel, L.~Kienle, B.~Wagner, and F.~Lofink, ``{Growth of Highly c-Axis Oriented AlScN Films on Commercial Substrates},'' {\em Micromachines}, vol.~13, no.~5, pp.~1--9, 2022.

\bibitem{Tang2023}
Z.~Tang, G.~Esteves, and R.~H. {Olsson III}, ``{Sub-quarter micrometer periodically poled Al$_{0.68}$Sc$_{0.32}$N for ultra-wideband photonics and acoustic devices},'' {\em Journal of Applied Physics}, vol.~134, no.~11, p.~114101, 2023.

\bibitem{ZhuW2021}
W.~Zhu, J.~Hayden, F.~He, J.~I. Yang, P.~Tipsawat, M.~D. Hossain, J.~P. Maria, and S.~Trolier-McKinstry, ``{Strongly temperature dependent ferroelectric switching in AlN, Al$_{1-x}$Sc$_{x}$N, and Al$_{1-x}$B$_{x}$N thin films},'' {\em Applied Physics Letters}, vol.~119, no.~6, 2021.

\end{thebibliography}

\end{document}